\documentclass[11pt]{article}
\usepackage{amsmath,amssymb}
\usepackage{float}
\usepackage{caption2}
\usepackage{graphicx}
\def\be{\begin{equation}}
\def\ee{\end{equation}}
\def\ba{\begin{eqnarray}}
\def\ea{\end{eqnarray}}

\begin{document}
\title{\textbf{\large{Effects of Multi-Field Phantom Inflation in Big Rip }}}
\vspace{1cm}
\author{IftikharAhmad\thanks{e-mail:dr.iftikhar@uog.edu.pk}, Farah
Naz\thanks{e-mail:farrahwarraich@uog.edu.pk}~~and~~Zaffar Iqbal\thanks{email:zaffar.iqbal@uog.edu.pk} \\
\textit{Institute of Physics and Mathematical Sciences}, \\
\textit{Department
of Mathematics, University of Gujrat}\\
\textit{ Pakistan.}}

\date{}
\maketitle
\begin{abstract}
In this paper we study the behavior of the multi-field in phantom
inflation, when massive scalar fields work collectively, in which
the scale factor is a power law. We evaluate its parameter values by
applying certain constraints on our model parameters, and we
investigate that before the Big Rip singularity occurs the universe
is in phantom inflationary phase. Furthermore, we calculate these
values for this period then compare with current observations of
CMB, BAO and observational Hubble data. We find that results may be
consistent with observations. This implies that in the dark-energy
equation of state (EOS) parameter $\omega_{DE}$ at the Big Rip
remains finite, with the divergence of pressure and dark energy
density.

\end{abstract}
\maketitle


\section{Introduction}

Cosmological observations made, by end of the last century and at
the beginning of this century, have conclusive evidence for late
cosmic acceleration \cite{S.J.P}. It is driven by an unknown fluid
violating strong energy condition (SEC), such that $\omega <
-\frac{1}{3}$, with $\omega$ being the ratio of pressure density to
energy density. This exotic fluid is known as dark energy
(DE). The phantom field, in which the parameter of the equation of
state $\omega < -1$, has still gained increasingly attention
\cite{R. R.C}, motivated by the study of dark energy, e.g.
\cite{CKW}, \cite{SSD}, \cite{HL}, \cite{NO}, \cite{GPZ},
\cite{ACL},\cite{KSS}, \cite{F}, \cite{BG}. The actions obeying the
Phantomlike behavior may be arise in supergravity \cite{N}, scalar
tensor gravity \cite{BEPS}, higher derivative gravity \cite{P},
braneworld \cite{SS}, string theory \cite{FR}, and other scenarios
\cite{CL,ST}, and also from quantum effects \cite{OW}. The visible
universe driven by the phantom field will evolve to a singularity,
in which the energy density become infinite at finite time, which is
called the Big Rip, see Refs. \cite{NOT5}, \cite{B}, \cite{S5},
\cite{BG22}, \cite{D23}, \cite{BGM24} for other future
singularities.\\

Recently, the little rip scenario has been proposed \cite{FLS}, in
which the current acceleration of universe is driven by the phantom
field, the energy density increases without bound, but $\omega$
tends to $-1$ asymptotically and rapidly, and thus the rip
singularity dose not occur within finite time, however, in little
rip scenario, the universe arrives at the singularity only at infinite time.\\

The simplest recognition of phantom field is a normal scalar field
with reverse sign in its dynamical term. This reverse sign results
in that, behave differently from the evolution of normal inflation
during the normal inflation, the phantom field during the phantom
inflation will be driven by its potential climb up along its
potential. With the use of General Relativity (GR)which based on
Friedmann equations, it observed that phantom-dominated epoch of the
universe goes faster, but ends up in the form of Big-Rip singularity
in a finite future time \cite{R. R.C}. Phantom dark energy fields
are characterized by violating the main energy condition, $\rho+p >
0$. Also the conservation equation
has the striking consequence that the energy density increases with
expansion and with condition $\omega< -1$ matter is called  Phantom
energy \cite{A.L}. On the basis of observational data Caldwell noted
that $EOS$ parameter $" \omega "$ has a very short range in the
neighborhood of $\omega = -1$ with more likelihood to the side of
$\omega<-1$. He found that this possibility could not be neglected
for the dark energy fluid. Alternatively,
a very good description of the evolution of universe, is discussed
in \cite{Q,R} and \cite{JQ}.

Here, we will argue that after a finite period of the Big Rip phase
the universe might return the evolution of observational universe.
However, it is possible that some times after the energy density of
the phantom field arrives at a high energy level, or before the rip
singularity of universe is arrived, the energy of field will be
released, and the universe reheats, after which the evolution of hot
"big bang" recurs. For phantom scalar field a power law cosmology is
defined by the cosmological scale factor evolving as $t^\beta$, see Refs. {\cite{0804}}.\\

In our model, when many massive fields work collectively to drive
phantom inflationary phase under certain constraints is known as
multi-field (Nfield) phantom cosmology, in which the scale factor is
a power law. In addition, it is very interesting to generalize the
above studies for Nfield phantom cosmology with various potentials
remains open.

Here, we study a general behavior of Nfield phantom inflation with
the scale factor given in terms of parameter $\beta$ with out any
dimension. The modified form of the scale factor is $
a(t)=a_{0}(({t_{s}-t})/({t_{s}-t_{0}}))^{\beta}$ in order to achieve
the self-stability, where $t_{s}$ is a required positive reference
time {\cite{Q}}.


The field starts from near an unstable equilibrium (taken to be at
the origin) and climb up the potential to a stable maxima. In the
phantom model, the observable Big Rip occurs during the climb up of
scalar field and its magnitude is at most of order $M_p$ the Planck
mass.

The arrangement of remaining sections of this paper is as; in
section $2$, we formulate the whole picture of phantom power-law
cosmology for multi-field. In section $3$ we consider observational
data to impose certain bounds to investigate results for the
multi-field parameters, and finally, section $4$ is devoted for
conclusion and discussion.

\section{ Nfield with Power-Law Expansion}

In this section we present Phantom cosmology under power law
expansion, when many fields are working collectively with $\phi_{i}$
is the $ith$ phantom scalar field. For the simplicity of our model,
we assume the homogenous and isotropic Friedmann-Robertson Walker
(FRW) background metric, \ba ds^2 = -dt^{2} +
a^{2}(t)\left[\frac{dr^{2}}{1-kr^{2}}+ r^{2}d{\Omega_2}^{2}\right].
\ea\\ Where $a(t)$ is a scale factor of the universe, $\Omega_2$ is
$2$-dimension unit sphere volume, $t$ is the cosmic time and $k$
represents the curvature of $3$-dimensional space with $k =1, 0, -1$
corresponds to open, flat and closed universe respectively.
Our model is given by the following action, see Refs. \cite{ALL,CX}:

\ba S = \int {d^{4}x}\sqrt{-g}\left[ \frac{R}{16\pi{G}}+
\sum_i\left[\frac{1}{2}{g^{\mu\nu}}{\partial_\mu{\phi_i}}{\partial_\nu{\phi_i}}
- V_i (\phi_i)\right] + L_{m}\right] \label{e1}.\ea

Which involves $N$ phantom scalar fields, where $R$ is the Ricci
scalar, $V_i $ potential of $ith$ phantom field and $G$ the Newton
gravitational constant. The Lagrangian $ L_{m}$ stands for the total
matter of the universe including (dark plus baryonic).
Finally, we concentrate on small redshifts, therefore, we are
neglected the radiation sector, with speed of light as unity
\cite{ALL}.\\

We assume the flat geometry of the universe i.e. $k=0$, for this
model, the $ith$ phantom field satisfies the equations,  \ba
H^{2}=\frac{1}{3M^2_p}{\sum_i}[\rho_{\phi_{i}}+\rho_{m_i}]{\label{e24}},\ea
\ba \ddot{\phi_i}+3H \dot{\phi_i}-V'(\phi_i)=0
{\label{e24a}},\ea\\
where $H(t)={\dot{a}}/{a}$ is the Hubble parameter  represents the
expansion rate of the universe at time $t$ and $M_p=(8\pi
G)^{-\frac{1}{2}}$ is the Planck mass. $H^2>0$ requires that in all
case for phantom evolution $\dot{\phi}^2$ must be smaller than its
potential energy. In the above expression $\rho_{\phi_i}$ and
$p_{\phi_i}$ are the energy density and pressure of the $ith$
phantom scalar field respectively, which are
 different from normal inflations model, reads
\cite{IPQ,IPQ1},

\ba
\rho_{\phi_{i}}=-\frac{1}{2}{\dot{\phi_{i}}}^{2}+V_{i}(\phi_{i})\label{e26}.\ea

\ba p_{\phi_{i}}=-\frac{1}{2}{\dot{\phi_{i}}}^{2}-
V_{i}(\phi_{i})\label{e27}.\ea By varying with respect to scalar, we
obtain the evaluation equation:

\ba \dot{\rho_{\phi_{i}}}+3H
 (p_{\phi_{i}}+\rho_{\phi{i}})=0\label{e28}.\ea
After simplification Eq. (\ref{e28}) can be evaluated for
multi-field as follows:
\ba\sum_i{\ddot\phi_{i}}+3H\sum_i{\dot\phi_{i}}
 -\sum_i\frac{dV_{i}}{d\phi_{i}}=0\label{e29}.\ea\\
As in phantom cosmology the dark energy sector is attributed to the
phantom fields, and thus its equation-of-state parameter is given by

\ba\omega_{DE}
=\frac{p_{DE}}{\rho_{DE}}=\frac{p_{\phi}}{\rho_{\phi}}\label{e001}.\ea
And for matter-dominated universe, we have expression
 \ba\omega_{m_i}
=\frac{p_{m_i}}{\rho_{m_i}}\label{e00}.\ea
Finally, in the case of matter density, Eq. (\ref{e28}) becomes \ba
\dot{\rho_{m_i}}+3H(1+\omega_{m_i})\rho_{m_i}=0\label{e01},\ea
\\with the simple form of its solution for multi-field is \ba
\sum_i\rho_{m_i}=\rho_{m_0}\sum_i({\frac{1}{a^{n_i}}})\label{e02},\ea
where $n_i=3(1+\omega_{m_i})$ and
here, we are dealing only with massive scalar fields, but the case
of massless scalar fields are neglected in this regime.
With the help of Eqs. (\ref{e26}) and (\ref{e27}), Eq. (\ref{e24a})
can calculate the result for Nfield as  \ba \dot
H=\frac{1}{6M^2_p}\left[3\sum_i{\dot\phi_{i}^{2}}-\sum_i{\rho_{m_i}}n_{i}\right]
\label{e30}.\ea

This result is consistent with single field inflation model when $N
= 1$, where $N$ stands for number of field. When we are working with
Phantom cosmology, we replace $t$ by $t_{s}-t$; the reference time
$t_{s}$ is sufficiently positive,
 then we obtain the scale factor by

\ba
a(t)=a_{0}{\left(\frac{t_{s}-t}{t_{s}-t_{0}}\right)}^{\beta}\label{e31},\ea
with the Hubble parameter and its derivatives with respect to time
is

\ba H(t)=\frac{-\beta}{(t_{s}-t)}\label{e34a},\ea

\ba\dot{H(t)}=-\frac{\beta}{(t_{s}-t)^{2}}\label{e34}.\ea

\begin{figure}[t]
\begin{center}
\includegraphics[width=8cm]{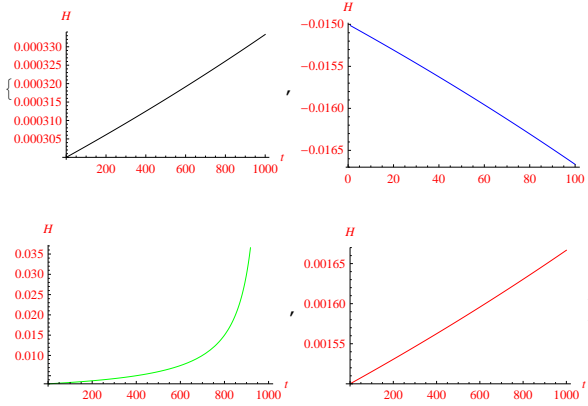}
\caption{ The evolution of $H$,  the upper-left sketch corresponds
to the parametrization of $(15)$ for $\beta=-3$, $t_s>t$, the
upper-right sketch corresponds to the parametrization of $(15)$ for
$\beta=15$, $t_s>t$,  the lower-left sketch corresponds to the
parametrization of $(15)$ for $\beta=-3$, $t_s=t$ and the
lower-right sketch corresponds to the parametrization of $(15)$ for
$\beta=-15$, $t_s>t$,
 }\label{xx}
\end{center}
\end{figure}

Now we investigate the behavior of universe which is depending on
the value of $\beta$, thus for the value of $\beta$ less than zero,
we observe an accelerating $(\ddot{a}(t)>0)$ universe and expanding
$(\dot{a}(t)>0)$ universe, we find that $\dot{H(t)}$ is positive
therein, which implies that it provides super acceleration, this is
only possible for phantom power-law cosmology case. In addition, for
the exponent $ \beta < 0$, and at late time $t = t_{s}$, as shown in
Figure $1$, the scale factor $a(t)$ and Hubble parameter $H(t)$ of
the universe both diverge
as a result it goes to a Big Rip.\\
Such actions are common in phantom cosmology and their realization
is a self-consistency test of our work, however, the important point
which is already discussed in Ref. {\cite{CBE}}.\\\\
By using Eqs. (\ref{e24}) and (\ref{e26}) in Eq. (\ref{e30}), we
find the potential $\langle V(t)\rangle=(\sum_i V_{i})/N$ which is
the average value of $V_i(t)$ . Since $ n_i=3(1+\omega_{m_i});\;\
1\leq i\leq N$. For dust dominated universe $\omega_{mi}\rightarrow
0 $, this implies that $n_i=3$ and as a result we obtain $\sum_i
n_{i}=3N$. Therefore, it is given by

\ba \langle
V(t)\rangle=\frac{\sum_iV_{i}(t)}{N}=\left[M_{p}^{2}\left(\frac{3\beta^{2}-\beta}
{(t_{s}-t)^{2}}\right)-\frac{5}{6}\frac{\rho_{m_0}}{{a_{0}}^{3}}
\left(\frac{t_{s}-t_{0}}{t_{s}-t}\right)^{3{\beta}}\right]\label{e36}.\ea

From Eq. (\ref{e30}) we can obtain

\ba\langle
{{\dot{\phi}}(t)}^2\rangle=\left[\frac{-2\beta{M_{p}^{2}}}
{(t_{s}-t)^{2}}+\frac{\rho_{m_0}}{{a_{0}}^{3}}
\left(\frac{t_{s}-t_{0}}{t_{s}-t}\right)^{3{\beta}}\right]\label{e37}.\ea

Using the values from Eqs. (\ref{e36}) and (\ref{e37}) in Eq.
(\ref{e26})

\ba \langle
\rho_{\phi}\rangle=\left[M_{p}^{2}{\frac{3\beta^{2}}{(t_{s}-t)^{2}}}-\frac{4\rho_{m_0}}
{3{a_{0}}^{3}}\left(\frac{t_{s}-t_{0}}{t_{s}-t}\right)^{3{\beta}}\right]\label{e38}.\ea

Again putting  Eqs. (\ref{e36}) and (\ref{e37}) into Eq. (\ref{e27})
we get

\ba \langle p_{\phi}\rangle=
\left[M_{p}^{2}\frac{(-3\beta^{2}-\beta+3\beta)}{(t_{s}-t)^{2}}+
(t_{s}-t)H(t)+\frac{\rho_{m0}}
{3{a_{0}}^{3}}\left(\frac{t_s-t_{0}}{t_s-t}\right)^{3{\beta}}\right]\label{e39}.\ea\\

In the case of phantom Nfield, the dark energy equation-of-state
parameter is

$$\omega_{DE}(t)={p_{\phi}}/{\rho_{\phi}},$$ which implies that

\ba w_{DE}(t)=(-1+{\frac{1}{\beta}})\label{e40}.\ea\\
We see that for Big Rip behavior, $\omega_{DE}(t)$ always having
finite value {\cite{NOT5}}.\\
For $\beta $ less than zero possesses additionally a positive
$\dot{H(t)}$  that leads to super-acceleration {\cite{S.Das}}. So
such kind of scenario expansion is always came with acceleration.
From Eq. (\ref{e40}), for $\omega < 0$ we come to know that the value of
$\omega$ is very narrow to phantom divide.\\
Furthermore, with the value of $\beta $ less than zero, at $t$ equal
to $t_{s}$ the scale factor and the Hubble parameter diverge, that
is the universe results to a Big Rip. We investigate that these
behaviors of Nfield phantom cosmology are very similar with result
of single field phantom cosmology with power-law {\cite{CBE}. Here,
we have limited the parameter $\beta $ small, however, it is
interesting to consider the phenomena of $ \beta\gg1$, i.e. there is
a new step, by which the density of dark energy observed might be
linked to that of inflation, as in the eternal expanding cyclic
scenario.\\

\section{\textbf{A Bound for $N$ for Nfield and  Fit the Observational Data}}
In this section, we apply the techniques that conform the values of
multi-field with the observational data. Now we are fitting the
observational data, presenting our results in the case of many
fields are working collectively.
We observed that our values for all parameters in {\textbf{Table I}}
are best fitted with minor error of accuracy, which is negligible
for large scale, we also provide the $1 \sigma$ bound of every
parameter. Similarly in {\textbf{Table II}} we present the maximum
possibility of the values up to  $1 \sigma$ bound for the derived
parameters, namely the power-law exponent $\beta$, the present
matter energy density value $\rho_{m_0}$, the present critical
energy density value $\rho_{c_0}$ and the Big Rip time $t_{s}$. As
we observed that $\beta$ is always less than zero, as
expected in consistent phantom cosmology.\\\\
Furthermore, we observed that the Big Rip time is one order of
magnitude greater than the present age of the universe, which shows
that such an outcome is predictable in phantom cosmology, unless one
include additional mechanisms as shown in Ref. {\cite{A19}.\\
\ba \langle V(t)\rangle\simeq
\left[{\frac{6.5\times10^{27}}{{(3.30\times10^{18}-t})^2}
\large}-2.52\times10^{-371}\times(3.30\times10^{18}-t)^{19.54}\right]\label{e42}.\ea

While when we consider $WMAP7$ data alone, it provides

\ba \langle V(t)\rangle \simeq
\left[{\frac{6.41\times10^{27}}{{(3.30\times10^{18}-t})^2}\Large}-1.98
\times10^{-369}\times(3.30\times10^{18}-t)^{19.37}\right]\label{e43}.\ea

Here we noted that in above results although the second term is very
small at early times of the universe, but becomes very important at
late times, this situation is close to the Big Rip. Now the scalar
field evolution at late time $(t\rightarrow t_{s})$, $\rho_{m_0}$
can be neglected and also set $a_0=1$.\\ Thus we obtain the new
results for Nfield in phantom cosmology as follows:

\ba {\dot{\phi}}=
\sqrt{\sum_i(\dot{\phi}_i)^2}\simeq{\frac{\sqrt{-2N\beta}{M_p}}
{(t_s-t)}}\label{e45b},\ea which implies that
\ba
\phi(t)\simeq{\sqrt{-2N\beta}{M_p}}\ln|{t_{s}-t}|\label{e46}.\ea\\
Thus the result is similar with single field Phantom model when $N =
1$ see {\cite{CBE}. However for large field we can find some new
interesting results for future. Additionally, the total change of
all fields is determined by the radial motion in field space ( for
example see Ref. \cite{IPQ1}), we have \ba
\Delta\phi\simeq{\frac{\sqrt{\sum_i ({\dot{\phi}_i})^2}}{H}}\simeq
-\sqrt{{\frac{-2N}{\beta}}}M_p\label{e40a}.\ea\\
Thus we see that for $\beta<-1$, the total absolute change of all
fields is directly proportional to the square root of $N$ and
inversely proportional to square root value of exponent $\beta$, for
the value $\beta=0$, this change is undefined, which is not possible
in Phantom cosmology.\\ Furthermore, for the combined
$WMAP7+BAO+H_{0}$, which implies that \ba
\phi(t)\cong{{\sqrt{N}}}[({-2.645\times10^{13}}).\ln|{3.30\times10^{18}-t}|]\label{e47},\ea
while single $WMAP7$ data provides that

\ba
\phi(t)\cong{{\sqrt{N}}}[({-2.643\times10^{13}}).\ln|{3.30\times10^{18}-t}|]\label{e48}.\ea
According to our exactions the phantom field and the kinetic energy
diverge at the Big Rip time.
$$\begin{tabular}{|c|c|c|c}
  \hline
  \hline
  \;\;Parameter\;\; & $H_{0}+WMAP7+BAO$ & \;\;WMAP7\;\;  \\\hline \hline
  $t_{0}$ & $13.78 \pm 0.11 Gyr [(4.33 \pm 0.04)\times 10^{17} sec]$&$13.71
  \pm 0.13 Gyr [(4.32 \pm 0.04)\times 10^{17} sec]$\\
  $H_{0}$ & $70.2_{+{1.3}}^{-1.4} \;\;\;km/s/Mpc$ & $71.4\pm 2.5 \;\;\;km/s/Mpc$\\
  $\Omega_{b_0}$ & $0.0455 \pm 0.0016$ & $0.0445 \pm 0.0028$\\
 $\Omega_{CDM_0}$ & $0.227 \pm 0.014$ & $0.217 \pm 0.026$\\
  \hline
  \hline
\end{tabular}$$

TABLE I: Observational maximum likelihood values in $1\sigma$
confidence level has taken from {\cite{1001}.

$$\begin{tabular}{|c|c|c|c}
  \hline
  \hline
  \;\;Parameter\;\; & $H_{0}+WMAP7+BAO$  & \;\;WMAP7\;\;  \\\hline \hline
  $\beta$ & $-6.52_{+ {0.24}}^{- 0.25}$ & $-6.51 \pm 0.4$ \\
  $\rho_{m_0}$ & $(-6.532 \pm 0.38) \times 10^{-27} \;\;\;kg/m^{3}$ &
  $(2.513\pm 0.27) \times 10^{-27}~~ kg/m^{3}$ \\
  $ \rho_{c_0}$ & $(9.4_{+{0.3}}^{-0.4}) \times 10^{-27}~~ kg/m^{3}$ &
  $(9.63 \pm 0.58) \times 10^{-27} ~~kg/m^{3}$\\
 $t_{s}$ & $104.83_{+{2.1}}^{-1.8}~ Gyr~~ \left[(3.30 \pm 0.06) \times 10^{18} sec\right]$ &
  $102.8 \pm 3.4 ~Gyr ~\left[(3.24 \pm 0.21) \times10^{18} sec\right]$ \\
  \hline
  \hline
\end{tabular}$$

TABLE II: Derived maximum likelihood values in $1\sigma$ confidence
level for the power-law exponent, present value of critical energy
density, present matter density, and $t_s$ at Big Rip time.

\section{Conclusion and Discussion }

In Big Rip phase, $\omega< -1$, the energy density of the all
phantom field together will increase with time, and arrive at a high
energy scale at finite time, and at this epoch the phantom field has
$\omega\simeq-1$. Here, we actually required that before
$\omega\simeq-1$, the phantom field must have arrived at a high
energy regime, which assures the occurrence of inflation. However,
the some Phantom fields loose their energy and jump back, this
process is continuous and inflation never goes to end, such type is
known as eternal phantom inflation which will be shown in later work.\\\\
In this paper, we study the Nfield phantom model, in which
collection of massive scalar fields drive it in early time of
universe. Now from Eq. (\ref{e47}) and Eq. (\ref{e48}), we find the
different values of $t$ and put them in Eq. (\ref{e42}) and Eq.
(\ref{e43}), to evaluate the average value of potential for the
Nfield Phantom power-law  i.e. $\langle V(\phi)\rangle$.

Thus, for the $WMAP7+BAO+H_{0}$ , the potential is fitted as  $$
\langle
V(\phi)\rangle\approx[{6.48\times10^{27}}e^{0.77\times10^{-13}({\phi}/\sqrt{N})}
-2.52\times10^{-371}e^{-7.44\times10^{-13}({\phi}/\sqrt{N})}],$$\\
and while for $WMAP7$, we can find
$$\langle V(\phi)\rangle\approx[{6.39\times10^{27}}e^{0.78\times10^{-13}({\phi}/\sqrt{N})}
-1.98\times10^{-369}e^{-7.5\times10^{-13}({\phi}/\sqrt{N})}],$$
respectively.

In our study of multi-field phantom cosmology, we observed that the
cosmic scale factor $a(t)$ is obeying the power law, for $N=1$, it
will give the result similar to single field, but when $N$ is very
very large, the value of $V(\phi)$ vary between $ 4\times 10^{27}$
to  $ 5.41\times 10^{27}$. When we construct the whole scenario, we
fit the observationally data of $WMAP7+BAO+H_{0}$ and $WMAP7$ alone
by applying bound on the multi-field by focusing on exponent $\beta$
and the Big Rip time $t_{s}$. By using separately WMAP7 data, we
obtained the value $\beta\cong {-6.523\pm 0.38}$, while the Big Rip
is observed at
$t_{s}\cong 102.8\pm 3.475\;Gyr$.\\
However, the dark energy equation of state parameter $\omega_{DE}$
lies below the phantom divide, it was expected and at Big Rip time
it always remains finite and nearly equal to $-1.1533$. Although the
phantom dark anergy density and pressure are diverging behavior at
the Big Rip. By using $ WMAP7+BAO+H_{0}$ data set alone we find
$\beta\approx {-6.51_{+ {0.24}}^{- 0.25}}$, while the Big Rip is
observed at $t_{s}\approx104.5_{+{1.9}}^{-2.0}~ Gyr $, in $1\sigma$
confidence level. Definitely, the subject of Nfield quantization of
such scenarios is open and needs further investigation on this.

\section{Acknowledgment}
We thank Yun-Song Piao for useful discussions and comments.We also
thank HEC (Higher Education Commission Pakistan) for providing the
facilities to carry out the research work.

\end{document}